\documentclass[epj]{svjour}
\usepackage{graphicx}

\let\I\i
\def\i{\mathrm{i}}
\def\e{\mathrm{e}}
\def\d{\mathrm{d}}
\def\half{{\textstyle{1\over2}}}

\def\h{{\scriptscriptstyle{1\over2}}}

\def\vec#1{\mbox{\boldmath$#1$}}
\def\svec#1{\mbox{{\scriptsize \boldmath$#1$}}}
\def\CG#1#2#3#4#5#6{C^{#5#6}_{#1#2#3#4}}

\begin{document}

\title{Pion electro-production in the Roper region 
in chiral quark models}

\author{%
B.~Golli\inst{1}
\and
S.~\v{S}irca\inst{2}
\and
M.~Fiolhais\inst{3}}

\institute{%
Faculty of Education,
              University of Ljubljana and J.~Stefan Institute,
              1000 Ljubljana, Slovenia
\and
Faculty of Mathematics and Physics,
              University of Ljubljana and J.~Stefan Institute,
              1000 Ljubljana, Slovenia
\and
Department of Physics and Centre for Computational Physics,
                  University of Coimbra,
                  3004-516 Coimbra, Portugal}
\date{\today}

\abstract{%
We present a method to calculate pion electro-production amplitudes 
in a coupled-channel framework incorporating quasi-bound quark-model 
states. The method offers a clear prescription how to extract the 
resonant part of the amplitudes, even in the presence of different 
decay channels and a strong mixing of neighbouring resonances. 
The method is applied to the calculation of the $M_{1-}$ and the 
$S_{1-}$ amplitudes in the P11 partial wave in a simple chiral 
quark model. A good agreement with the observed $M_{1-}$ amplitude 
is found with a significant contribution from the pion cloud.
The same effect is also prominent in the $S_{1-}$ amplitude but a 
rather uncertain data prevent us to draw a definitive conclusion. 
}


\maketitle

\section{\label{sec:}Introduction}

The study of baryon resonances above the inelastic 
threshold is frequently characterized by a strong 
interplay of different decay channels, mixing of 
different model states with equal quantum numbers,
as well as the presence of background processes.
These effects can almost completely obscure the relevant 
information about the resonance under investigation and 
make it hard to establish 
a clear connection of 
the data extracted in meson scattering and electro-weak 
processes to the properties obtained in model calculations.
While these effects influence only little the properties 
of the lowest resonance $\Delta(1232)$, they are strongly 
present already in the case of the $N(1440)$ (Roper) 
resonance.

In our previous work \cite{EPJ2008} we have developed 
a general method to incorporate excited baryons represented 
as quasi-bound quark-model states into a coupled channel 
calculation of pion scattering using the $K$-matrix approach.
The method ensures unitarity through the symmetry of 
the $K$ matrix as well as the proper asymptotic conditions.
We were able to explain a rather intriguing behaviour of the 
scattering amplitudes in the region of the Roper resonance 
through the inclusion of the $\pi\Delta$ and $\sigma N$  
inelastic channels.
In this work we extend the formalism to the calculation 
of electro-production amplitudes.

The electromagnetic properties of the Roper resonance have 
been studied in several models involving quark, me\-sonic 
and/or gluonic degrees of freedom focusing mainly on the 
calculation of the electro-excitation part of the process.
The constituent quark model, assuming a (1s)$^2$(2s)$^1$
configuration, does not yield sensible results, even predicting 
the wrong sign of the $A_{1/2}$ amplitude at the photon point.
It has been suggested that additional degrees of freedom, such as 
explicit excitations of the gluon field~\cite{Li}, the glueball 
field~\cite{broRPA,PLB2001}, or chiral fields~\cite{Krehl}
may be relevant for the formation of the Roper resonance.
The importance of a correct relativistic treatment in 
constituent quark models has been emphasized
in~\cite{Capstick,Weber,Cardarelli,Julia-Diaz04,Inna07}
yielding the correct sign at the photon point along with 
the sign change of $A_{1/2}$ at  $Q^2\sim 0.5$~(GeV$/c)^2$.
In order to reproduce its relatively large value at
the photon point, the effects of the meson cloud have
to be included as indicated in \cite{Cano,Dong,Tiator88}.
The quark charge densities inducing the nucleon to Roper
transition have been determined from the phenomenological
analysis \cite{Tiator09} confirming the existence of a narrow 
central region and a broad outer band.

The present work is -- to the best of our knowledge -- the 
first attempt to apply a quark model of baryons to calculate 
full electro-production amplitudes in the case  of a strong 
background  and in the presence of open inelastic channels.
Such processes are usually calculated in models based on baryon
and meson degrees of freedom which involve numerous adjustable
parameters, and in which resonances are incorporated in distinct ways.
In the effective-Lagrangian unitary isobar model MAID \cite{MAID} 
the electro-magnetic resonant vertices are dressed (i.e. they
already contain the meson-cloud contributions) while in the dynamical
models, for example, DMT \cite{DMT} or SL \cite{SL}, the resonances
are bare and the meson-cloud effects are generated dynamically.
The dynamical models were put to thorough scrutiny when faced with
a large body of data coming from recent $N\to\Delta$ experiments,
but only preliminary results exist in the Roper region 
\cite{Julia-Diaz09}.

In the next section we briefly review the construction
of meson-baryon channel states which incorporate the 
quasi-bound quark-model states corresponding to the nucleon 
and its higher resonances.
The construction of the multi-channel $K$ matrix is discussed
and the method to solve the Lippmann-Schwinger equation
for the meson amplitudes is outlined.

In sec.~\ref{sec:production} we give the formulas
for the matrix elements of the multi-channel $K$ matrix 
involving the photon-baryon channel, and the relation
to the pion electro-pro\-duction amplitude is established.

In sec.~\ref{sec:resonance} we discuss the form of
the production amplitude close to a resonance and 
explain how the resonant part of the amplitude 
can be isolated.
Furthermore, we discuss the origin of different terms
contributing to the background part of the amplitude.

The multipole expansion for the P11 wave is introduced in 
sec.~\ref{sec:multipoles} and the extraction of the helicity 
amplitudes from the corresponding electro-production 
amplitude is explained in sec.~\ref{sec:excitation}.
We show that our method yields the standard relation
between these two amplitudes and determines
the relative sign between them.

We calculate the electro-production amplitudes 
in the region of the Roper resonance using the Cloudy 
Bag Model with the same set of parameters as used in the 
calculation of scattering amplitudes in our previous work.
The results are presented and discussed in sec.~\ref{sec:results}
and summarized in the last section.

\section{\label{sec:formalism}Incorporating quark-model states 
into a multi-channel formalism}

We consider a class of chiral quark models in which mesons
couple linearly to the quark core:
\begin{eqnarray}
 H' &=& 
  \int\d k \sum_{lmt}\left\{\omega_k\,a^\dagger_{lmt}(k)a_{lmt}(k)
\right. \nonumber\\ && \left.
    + \left[V_{lmt}(k) a_{lmt}(k) 
        + V_{lmt}^\dagger(k)\,a^\dagger_{lmt}(k)\right] \right\},
\label{Hpi}
\end{eqnarray}
where $a^\dagger_{lmt}(k)$ is the creation operator for a meson 
with angular momentum $l$ and the 
third components of spin $m$ and isospin $t$.
If we include only  $l=1$ pions, the form of the source is
\begin{equation}
      V^\pi_{mt}(k) = -v(k)\sum_{i=1}^3 \sigma_m^i\tau_t^i\,.
\label{Vmt}
\end{equation}
The quark operator $V_{lmt}(k)$ depends on the model and
includes the possibility that the quarks change their
radial function which is specified by the reduced matrix elements
$V_{BB'}(k)=\langle B||V(k)||B'\rangle$, where $B$ are the bare
baryon states (e.g. the bare nucleon, $\Delta$, Roper, $\ldots$)

We have shown that in such models
it is possible to find an exact expression for the $K$ matrix
without explicitly specifying the form of the asymptotic states.
In the basis with good total angular momentum $J$ and 
isospin $I$, the elements of the $K$~matrix take the form
\begin{eqnarray}
   K_{MB\,M'B'}^{JI} &=&  -\pi\mathcal{N}_{MB}
     \langle\Psi^{M'B'}_{JI}||V^{M}(k)||\widetilde{\Psi}_{B}\rangle\,,
\nonumber\\
   \mathcal{N}_{MB} &=& \sqrt{\omega_M E_{B} \over k_M W}\,,
\quad  E_B(k)=\sqrt{{M_B}^2+k^2}\,,
\label{KHHp}
\end{eqnarray}
where  $\omega_M$ and $k_M$ are the energy and momentum of the meson.
Here $\Psi_{JI}^{MB}$ is the principal-value state corresponding to
the channel specified by the meson $M$ ($\pi$, $\sigma$, $\ldots$)
and the baryon $B$ ($N$, $\Delta$, $\ldots$):
\begin{eqnarray}
|\Psi_{JI}^{MB}\rangle &=& \mathcal{N}_{MB}\left\{
   \sum_{\mathcal{R}}c_{\mathcal{R}}^{MB}|\Phi_{\mathcal{R}}\rangle
   + [a^\dagger(k_M)|\widetilde{\Psi}_B\rangle]^{JI} 
\right. \nonumber\\ && \left. 
+ \sum_{M'B'}
   \int {\d k\>
       \chi_{JI}^{M'B'\,MB}(k)\over\omega_k+E_{B'}(k)-W}\,
      [a^\dagger(k)|\widetilde{\Psi}_{B'}\rangle]^{JI}\right\},\kern12pt
\label{PsiH}
\end{eqnarray}
normalized as
\begin{eqnarray}
\langle \Psi_{J'I'}^{M'B'}(W')|\Psi_{JI}^{MB}(W)\rangle
&=& \delta_{J'J}\delta_{I'I}\left(\delta_{M'B',MB}
 + \mathbf{K}^2_{M'B'\,MB}\right)
\nonumber\\ && \times
\delta(W-W').
\label{normPsiH}
\end{eqnarray}
where $W$ is the invariant energy of the system.
The first term is the sum over {\em bare\/} three-quark states 
$\Phi_{\mathcal{R}}$ involving different excitations of the quark 
core, the next term, which defines the channel, corresponds 
to the free meson and the baryon, and the third term 
represents meson clouds around different isobars.
The sum in the latter term includes also inelastic channels in 
which case the integration over the mass of the unstable intermediate 
hadrons ($\sigma$-meson, $\Delta$, $\ldots$) is implied.
The state $\widetilde{\Psi}_{B}$ in (\ref{KHHp}) and 
(\ref{PsiH}) represents either the nucleon or the intermediate 
isobar decaying into the nucleon and the pion.
In the latter case the state is described by (\ref{PsiH})
with $MB=\pi N$ and normalized to
$\delta_{J'J}\delta_{I'I}\delta_{M'B', MB}\delta(M_B-M_B')$ 
(instead of (\ref{normPsiH})), where $W$ has been replaced by  
the invariant mass of the $\pi N$ system $M_B$.
The on-shell meson amplitudes $\chi_{JI}^{M'B'\,MB}$ are proportional 
to the corresponding matrix elements of the on-shell $K$ matrix
\begin{equation}
   K_{M'B'\,MB}  = \pi\,\mathcal{N}_{M'B'}\mathcal{N}_{MB}\,
             \chi_{JI}^{M'B'\,MB}(k_{M'}) \,.
\label{chi2K}
\end{equation}
From the variational principle for the $K$ matrix it is possible
to derive a set of integral equations for the $\chi$ amplitude which
is equivalent to the Lippmann-Schwinger equation for the $K$ matrix.
The resulting expression for $\chi$ can be written in the form
\begin{equation}
   \chi_{JI}^{M'B'\,MB}(k) 
     = -\sum_{\mathcal{R}}\widetilde{c}^{MB}_{\mathcal{R}}
    \widetilde{\cal V}^{M'}_{B'\mathcal{R}}(k)
       + \mathcal{D}_{JI}^{M'B'\,MB}(k)\,.
\label{sol4chi}
\end{equation}
The states $\Phi_\mathcal{R}$ are not eigenstates of the Hamiltonian 
and therefore mix:
$
   \widetilde{\Phi}_{\mathcal{R}} = \sum_{\mathcal{R}'} 
   u_{\mathcal{R}\mathcal{R}'} \Phi_{\mathcal{R}'}\,.
$
As a consequence,
\begin{equation}
    \widetilde{\cal V}^M_{B\mathcal{R}} = 
        \sum_{\mathcal{R}'}u_{\mathcal{R}\mathcal{R}'}
         {\cal V}^M_{B\mathcal{R}'}\,,
\quad
    \widetilde{c}_{\mathcal{R}}^{MB} = {\widetilde{\cal V}^M_{B\mathcal{R}}
             \over Z_{\mathcal{R}}(W) (W-M_{\mathcal{R}})}\,,
\label{calVmix}
\end{equation}
where $\mathcal{V}^M_{B\mathcal{R}}$ are the dressed matrix elements
of the quark-meson interaction between the resonant state and the
bary\-on state in channel $MB$, and $Z_{\mathcal{R}}$ is the 
wave-function normalization.

\section{\label{sec:production}The $\pi N$ electro-production amplitudes}

The EM interaction Hamiltonian is assumed to be of the form
$$
  H_\gamma ={1\over\sqrt{2\pi}^3}\int\d\vec{k}_\gamma  \sum_\mu\,
   \left[\tilde{V}^\gamma_\mu(\vec{k}_\gamma)a_\mu(\vec{k}_\gamma) 
                  + \hbox{h.c.}\right]\,,
$$
where  $\vec{k}_\gamma$ and $\mu$ are the momentum 
and the polarization of the incident photon, and
\begin{equation}
\tilde{V}^\gamma_\mu(\vec{k}_\gamma)
  = {\mathrm{e}_0\over\sqrt{2\omega_\gamma}}
    \int\d\vec{r}\,\vec{\varepsilon}_\mu\cdot\vec{j}(\vec{r})
    \,\e^{\i\svec{k}_\gamma\cdot\svec{r}}\,.
\label{Vgamma}
\end{equation}
The state representing the photon-nucleon system reads
\begin{equation}
|\Psi_N(m_s,m_t;\vec{k}_\gamma,\mu)\rangle
  = \,\mathcal{N}_\gamma
     a_\mu^\dagger(\vec{k}_\gamma)|\Psi_N(m_s m_t)\rangle\,,
\end{equation}
\begin{equation}
   \mathcal{N}_\gamma 
   = \sqrt{k_\gamma\omega_\gamma}\sqrt{E_N^\gamma\over W}\,.
\end{equation}
Here $m_s$ and $m_t$ are the third components of the 
nucleon spin  and isospin,
$
   \omega_\gamma = (W^2-M_N^2-Q^2)/2W\,,
$
$
   k_\gamma^2 = \omega_\gamma^2 + Q^2\,,
$
$
   E_N^\gamma = W-\omega_\gamma\,,
$
with $Q^2$ measuring the photon virtuality.
In the type of models we are considering here, the current and the 
charge density operators can be split into quark and pion parts:
\begin{eqnarray}
  \vec{j}(\vec{r})  &=&
  \bar{\psi}\vec{\gamma}({\textstyle{1\over6}} + \half\tau_0)\psi
  + \i \sum_t t \pi_t(\vec{r})\vec{\nabla}\pi_{-t}(\vec{r})\; ,
\label{current}\\
  \rho(\vec{r}) & = &
  \bar{\psi}\gamma_0({\textstyle{1\over6}} + \half\tau_0)\psi
  - \i \sum_t t \pi_t(\vec{r}) P^\pi_{-t}(\vec{r})\;,
\label{charge}
\end{eqnarray}
where $P^\pi$ stands for the canonically conjugate pion field.
The amplitude for pion electro-production on the nucleon
is defined as
\begin{equation}
\mathcal{M}^{JI}_{\pi N}
  = -{\mathcal{N}_\gamma\over \sqrt{k_0 k_\gamma}}\,
     \langle{\Psi}_{JI}^{(+)}(m_Jm_I;k_0,l)|
      \tilde{V}^\gamma_\mu(\vec{k}_\gamma)|\Psi_N(m_s m_t)
       \rangle\,.
\label{MNJI}
\end{equation}
It is related to the corresponding $T$ matrix through
 $T=\sqrt{k_0 k_\gamma/8\pi}\;\mathcal{M}$.
In (\ref{MNJI}) $m_J$ and $m_I$ are the third components of the 
spin and the isospin,  and $k_0$ is the outgoing pion momentum.

The $K$-matrix elements for electro-production corresponding to 
different channels $MB$ ($\pi N, \pi\Delta$, $\sigma N, \ldots$) 
are introduced similarly as in (\ref{MNJI}) by replacing the 
state $\Psi_{JI}^{(+)}$ by the principal-value state (\ref{MNJI}):
\begin{equation}
\mathcal{M}^{K\,{JI}}_{MB}
  = -{\mathcal{N}_\gamma\over \sqrt{k_0 k_\gamma}}\,
     \langle{\Psi}_{JI}^{MB}(m_Jm_I;k_0,l)|
      \tilde{V}^\gamma_\mu(\vec{k}_\gamma)|\Psi_N(m_s m_t)
  \rangle .
\label{MKNJI}
\end{equation}
They are related to the electro-production amplitudes\break 
through $\mathcal{M}=\mathcal{M}^K+\i\, T\mathcal{M}^K$.
(This trivially follows from the Heitler's equation
$T=K+\i\, TK$ since the proportionality factor between $T$ and 
$\mathcal{M}$ is the same as between $K$ and $\mathcal{M}^K$.) 
In principle, the equation for $\mathcal{M}$ involves also the 
matrix elements corresponding to Compton scattering, but they 
can be neglected since they are orders of magnitude smaller 
than those containing the strong interaction.
In the region of the Roper resonance in the P11 partial wave 
it suffices to  consider only the $\pi\Delta$ and the $\sigma N$ 
inelastic channels and the equation reads
\begin{eqnarray}
\mathcal{M}_{\pi N}(W) &=& \mathcal{M}_{\pi N}^K(W)
      +\i\biggl[T_{\pi N\pi N}(W)\mathcal{M}_{\pi N}^K(W)
\biggr. \nonumber \\ &+& \biggl.
       \int_{M_N+m_\pi}^{W-m_\pi} \kern-3pt\d M_\Delta \,
           T_{\pi N\pi\Delta}(W,M_\Delta)
        \mathcal{M}_{\pi\Delta}^K(W,M_\Delta)
\biggr. \nonumber \\ &+& \biggl.
       \int_{2m_\pi}^{W-M_N} \kern-3pt\d \mu 
          \,T_{\pi N\sigma N}(W,\mu)
            \mathcal{M}_{\sigma N}^K(W,\mu)
\biggr]\,.
\label{MgamNpiNa}
\end{eqnarray}
Here $M_\Delta$ denotes the invariant mass of the $\pi N$ system 
originating from the decaying $\Delta$ isobar and $\mu$ the invariant 
mass of the two-pion system from the decaying $\sigma$-meson.

\section{\label{sec:resonance}The behaviour of the amplitudes 
close to a resonance}

From (\ref{chi2K}), (\ref{sol4chi}) and (\ref{calVmix}) it 
follows that close to a resonance, denoted by $\mathcal{R}$, 
the $K$-matrix element between the elastic channel and an 
arbitrary channel $MB$ can be split in the resonant 
and the background parts
\begin{equation}
    K_{\pi N\, MB} = 
  -\pi\sqrt{\omega_0\omega_ME_NE_B\over k_0k_MW^2}
   \, \widetilde{c}_{\mathcal{R}}^{MB}
    \widetilde{\mathcal{V}}^\pi_{N{\mathcal{R}}}(k_0) 
     + K_{\pi N\,MB}^{\mathrm{bkg}}\,.
\label{Ksplit}
\end{equation}
Collecting the terms containing the coefficient 
$\widetilde{c}_{\mathcal{R}}^{MB}$ in (\ref{PsiH}) and
in (\ref{sol4chi}) and expressing it in terms of 
$ K_{\pi N\, MB}-K_{\pi N\,MB}^{\mathrm{bkg}}$ using (\ref{Ksplit}), 
the principal-value state (\ref{PsiH}) takes the form
\begin{equation}
   |\Psi^{MB}_{JI}\rangle =  
          -K_{\pi N\,MB} \sqrt{k_0W\over\pi^2\omega_0E_N}
           {\sqrt{\mathcal{Z}_{\mathcal{R}}}\over 
              \widetilde{\mathcal{V}}^\pi_{N{\mathcal{R}}}}\, 
          |\widehat{\Psi}^{\mathrm{res}}\rangle 
        +  |\Psi^{MB\,\mathrm{(bkg)}}_{JI}\rangle ,
\label{PsiH2PsiRes}
\end{equation}
where
\begin{eqnarray}
  |\widehat{\Psi}_{\mathcal{R}}^{\mathrm{res}}\rangle 
&=&
{1\over\sqrt{\mathcal{Z}_{\mathcal{R}}}}
 \Biggl\{|\Phi_{\mathcal{R}}\rangle
   - \int\d k\,{\widetilde{\mathcal{V}}^\pi_{N\mathcal{R}}(k)
      [a^\dagger(k)|\Psi_N\rangle]^{JI} \over\omega_k+E_N(k)-W}
\Biggr. \nonumber\\ && \Biggr.   
- \sum_{MB}\int\d k\,{\widetilde{\mathcal{V}}^M_{B\mathcal{R}}(k)
      [a^\dagger(k)|\widehat{\Psi}_B\rangle]^{JI} 
                                    \over\omega_k+E_B(k)-W}
\Biggr\}\,,
\label{PsiRes}
\end{eqnarray}
while $\Psi^{MB\,\mathrm{(bkg)}}_{JI}$ has the form of (\ref{PsiH})
without the terms containing $\widetilde{c}_{\mathcal{R}}^{MB}$
plus a term in the form of the resonant part of (\ref{PsiH2PsiRes})
in which $K_{\pi N\, MB}$ is replaced by $ K_{\pi N\,MB}^{\mathrm{bkg}}$.

We can now split also the amplitude  (\ref{MKNJI}) into 
the resonant and the background parts:
\begin{eqnarray}
  \mathcal{M}_{MB}^K &=& 
    \sqrt{\omega_\gamma E_N^\gamma\over\pi^2\omega_0E_N}
                    {\sqrt{\mathcal{Z}_{\mathcal{R}}}\over 
                     \mathcal{V}_{N{\mathcal{R}}}}\,
        K_{\pi N\,MB}\langle\widehat{\Psi}_{\mathcal{R}}^\mathrm{(res)}(W)|
          \tilde{V}^\gamma|\Psi_N\rangle
\nonumber\\
&& +   \mathcal{M}_{MB}^{K\,\mathrm{(bkg)}}           \,.
\label{MKgamNpiH}
\end{eqnarray}
We see that the resonant part depends on the channel indices only 
through the corresponding element of the scattering $K$ matrix. 
Next we plug (\ref{MKgamNpiH}) into (\ref{MgamNpiNa}) and take 
into account the relation between the $T$ and the $K$ matrices 
for scattering ($T=K+\i TK$).
The resonant part of the electro-production amplitudes then reads
\begin{equation}
\mathcal{M}_{\pi N}^\mathrm{(res)}  = 
    -\sqrt{\omega_\gamma E_N^\gamma\over\pi^2\omega_0E_N}
      {\sqrt{Z_{\mathcal{R}}}\over \mathcal{V}_{N{\mathcal{R}}}}
\,\langle\widehat{\Psi}_{\mathcal{R}}^\mathrm{(res)}(W)|\tilde{V}^\gamma
                 |\Psi_N\rangle\, T_{\pi N\pi N}\,,
\label{Mres}
\end{equation}
while the background part satisfies
\begin{eqnarray}
\mathcal{M}_{\pi N}^\mathrm{(bkg)} &=& 
       \mathcal{M}_{\pi N}^{K\,\mathrm{(bkg)}}
      +\i\biggl[T_{\pi N\pi N}\mathcal{M}_{\pi N}^{K\,\mathrm{(bkg)}} 
\biggr. \nonumber \\ && \biggl.
+ \overline{T}_{\pi N\pi\Delta}
  \overline{\mathcal{M}}_{\pi\Delta}^{K\,\mathrm{(bkg)}} 
+ \overline{T}_{\pi N\sigma N}
  \overline{\mathcal{M}}_{\sigma N}^{K\,\mathrm{(bkg)}}
\biggr],\kern24pt
\label{Mnon}
\end{eqnarray}
where $\overline{T}$ and $\overline{\mathcal{M}}$ are
the amplitudes averaged over the invariant masses of the
intermediate hadron using the averaging procedure introduced
in \cite{EPJ2008}.
The background part of (\ref{MKgamNpiH}) can be cast in the form
\begin{eqnarray}
  \mathcal{M}_{MB}^{K\,\mathrm{(bkg)}} &=& 
     \sqrt{\omega_\gamma E_N^\gamma \over\omega_0E_N}\,
    {K^\mathrm{(bkg)}_{\pi N\,MB}\over
             \pi\mathcal{V}^\pi_{N{\mathcal{R}}}(k_0)}\;
     \langle\widehat{\Psi}_{\mathcal{R}}^\mathrm{(res)}|\tilde{V}^\gamma
                                           |\Psi_N\rangle
\label{KHbg-res}\\ &+&  
  \sqrt{\omega_\gamma\omega_{M}E_N^\gamma E_B\over k_0k_{M}W^2}\,
  \biggl[
  \sum_{{\mathcal{R}}'\ne {\mathcal{R}}} {c_{\mathcal{R}'}^{MB}} 
          \langle\widehat{\Psi}_{\mathcal{R}'}
           |\tilde{V}^\gamma|\Psi_N \rangle\kern15pt 
  \biggr.
\label{KHoth-res}\\ &+&  
   \langle\widehat{\Psi}_{\mathcal{R}}^{MB\mathrm{(non)}}|\tilde{V}^\gamma
                                            |\Psi_N\rangle
\label{KHdir}\\ &+&
 \biggl.  
 \left[\langle\Psi_B|a(k_M)\right]^{JI}\tilde{V}^\gamma
                                                  |\Psi_N\rangle
   \biggr]\,.           
\label{KHa}
\end{eqnarray}
Here $\widehat{\Psi}_{\mathcal{R}'}$ corresponds to a resonance 
${\mathcal{R}'}$ other than the chosen one (i.e. $\mathcal{R}$) 
and  has the form (\ref{PsiRes}) with $\mathcal{R}$ replaced by 
${\mathcal{R}'}$; the corresponding matrix 
element does not depend on the channel indices $MB$.
In the P11 partial wave this type of contribution is dominated 
by the ground state in which case $\Phi_{\mathcal{R}'}$ 
is replaced by the exact ground state $\Psi_N$.
The state $\widehat{\Psi}_{\mathcal{R}}^{MB\mathrm{(non)}}$ 
in (\ref{KHdir}) originates from the non-resonant part of 
(\ref{sol4chi}); it has the form of (\ref{PsiRes}) without the
leading $\Phi_{\mathcal{R}}$ and with $\mathcal{D}_{JI}^{M'B'\,MB}(k)$ 
replacing $\mathcal{V}_{N{\mathcal{R}}}$.
The last term (\ref{KHa}) can be further manipulated by
commuting $a(k_0)$ through $\tilde{V}^\gamma$.  
From (\ref{Hpi}) it follows
$
    a_{mt}(k)|\Psi_N\rangle = 
      -V^\dagger_{mt}(k)(\omega_k + H - M_N)^{-1}|\Psi_N\rangle\,,
$
which yields (for $I=J$)
\begin{eqnarray}
&&\left[\langle\Psi_{J'=I'}|a(k)\right]^{JI=J}_{m_Jm_I}
\phantom|\tilde{V}\!\phantom|^{T0}_{L\mu}|\Psi_N m_s m_t\rangle =
\nonumber\\ 
&& - \sum_j g_{\h J'j}^{JLT}{\langle \Psi_{J'}||\!
         \phantom|\tilde{V}\phantom|^{T}_{L}||\Psi_j\rangle
    \langle \Psi_N||V(k)||\Psi_j\rangle
    \over
    \omega_k + E_j(k) - M_N}\,
    \CG{\h}{m_s}{L}{\mu}{J}{m_J}\CG{\h}{m_t}{T}{0}{J}{m_I}
\nonumber\\ 
&& +
  \langle\Psi_{J'}m_s'm_t'|[a_{mt}(k),\tilde{V}\!\phantom|^{T0}_{L\mu}]
   | \Psi_N m_s m_t\rangle
   \CG{J'}{m_s'}{1}{m}{J}{m_J}\CG{J'}{m_t'}{1}{t}{J}{m_I}\,.
\nonumber\\
\label{aVgamma}
\end{eqnarray}
Here $\tilde{V}\!\phantom|^{T0}_{L\mu}$ is a chosen multipole 
of the EM interaction (discussed in the next section)
where $T=0$ and 1 stand for the isoscalar and the isovector
part, respectively, and
$$
   g_{J''J'j}^{JLT} = (2J'+1)(2J''+1)W(LJ'J''1;jJ)W(TJ'J''1;jJ),
$$
where $W$ are the Racah coefficients.
The first term leads to a u-channel contribution with the 
intermediate states $\Psi_j$ dominated by the nucleon 
and the delta, while the second term corresponds to the pion 
pole term.

\section{\label{sec:multipoles}Multipole decomposition}

Expanding (\ref{Vgamma}) into multipoles, we have for the 
$M_{1-}$ and the $S_{1-}$ amplitudes: 
\begin{eqnarray}
M_{1-}^{(1/2)} &=& -\sqrt{\omega_\gamma E_N^\gamma\over6 k_0 W}
\langle \Psi_{JI}^{(+)}||\tilde{V}^{M1}_{(T=1)}||\Psi_N\rangle\,,
\nonumber\\
M_{1-}^{(0)} &=& -\sqrt{\omega_\gamma E_N^\gamma\over18 k_0 W}
\langle \Psi_{JI}^{(+)}||\tilde{V}^{M1}_{(T=0)}||\Psi_N\rangle\,,
\label{defM1-}
\end{eqnarray}
\begin{eqnarray}
S_{1-}^{(1/2)} &=& -\sqrt{\omega_\gamma E_N^\gamma\over2 k_0 W}
\langle \Psi_{JI}^{(+)}||\tilde{V}^{C0}_{(T=1)}||\Psi_N\rangle\,,
\nonumber\\
S_{1-}^{(0)} &=& -\sqrt{\omega_\gamma E_N^\gamma\over6 k_0 W}
\langle \Psi_{JI}^{(+)}||\tilde{V}^{C0}_{(T=0)}||\Psi_N\rangle\,,
\label{defS1-}
\end{eqnarray}
related to $\pi^0$ production amplitude on the proton and neutron as
\begin{equation}
{}_{p,n}M_{1-}^{(1/2)} = M_{1-}^{(0)} \pm {1\over3}\, M_{1-}^{(1/2)}\,,
\label{Mpn}
\end{equation}
and analogously for the $S_{1-}$.
Here 
\begin{equation}
 \tilde{V}^{C0}(k_\gamma)=  \sqrt{4\pi\alpha\over2\omega_\gamma}\,
   \int\d\vec{r}\,\rho_\mathrm{EM}\, j_0(k_\gamma r) 
\end{equation}
is the Coulomb multipole. 
The same formulas apply to the $\mathcal{M}^K$ amplitudes
that enter (\ref{Mres}) and  (\ref{Mnon}) provided
$\Psi_{JI}^{(+)}$ is replaced by (\ref{PsiH}).

\section{\label{sec:excitation}Helicity amplitudes}

At the resonant energy ($W=M_{\mathcal{R}}$) the transition 
amplitude appearing in (\ref{Mres}) between the ground state 
and the resonant state $\widehat{\Psi}_{\mathcal{R}}^{\mathrm{res}}$ 
corresponds to the helicity amplitude for electro-excitation 
of the resonance.
While the sign of the electro-production amplitudes is
fixed by (\ref{defM1-}) and (\ref{defS1-}), the sign of 
the helicity amplitude (as well as the sign of the pion 
decay amplitude) depends on the relative phase between the 
wave functions of the excited state and the ground state.
The helicity amplitudes for the Roper resonance are defined 
\cite{Inna07}, \cite{MAID}  as 
\begin{eqnarray}
A_{1/2} &=& -\xi_{\mathcal{R}} \,
            \langle\widehat{\Psi}_{\mathcal{R}}^{\mathrm{res}}\,  
         (m_s'=\half)|\tilde{V}^{M1}|\Psi_N\,(m_s=-\half)\rangle\,,
\kern12pt\label{Ah}\\
S_{1/2} &=&  -\xi_{\mathcal{R}} \, 
            \langle \widehat{\Psi}_{\mathcal{R}}^{\mathrm{res}}\, 
           (m_s'=\half)|\tilde{V}^{C0}|\Psi_N\,(m_s=\half)\rangle\,.
\label{Sh}
\end{eqnarray}
Here $\xi_{\mathcal{R}} = {\rm sign}(g_{\pi N\mathcal{R}}/g_{\pi NN})$.
Since $\omega_\gamma$ does not enter the final expression for 
the production amplitudes, it has been adopted to use the value 
$\omega_\gamma(Q^2=0)\equiv k_W$ in the denominator of (\ref{Vgamma}) 
and in the numerator of (\ref{defM1-}) and(\ref{defS1-}) also in 
the region $Q^2\ne 0$.

We now show that our formalism yields the familiar relation
between the electro-production and helicity amplitudes.
From (\ref{Mres}), (\ref{defM1-}) and (\ref{Mpn}) it follows
\begin{eqnarray}
{\rm Im}\,{}_pM_{1-}^{(1/2)} &=& 
     -{1\over3}\sqrt{k_W E_N^\gamma\over6\pi^2\omega_0E_N}
      {\sqrt{Z_{\mathcal{R}}}\over \mathcal{V}^\pi_{N\mathcal{R}}}
\;{\rm Im} T_{\pi N\pi N}
\left(-{3\over\sqrt2}\right)
\nonumber\\
&&\times
\langle \widehat{\Psi}_{\mathcal{R}}^\mathrm{(res)}\,(m_s=\half)|
\tilde{V}^{M1}|\Psi_N\, (m_s=-\half) \rangle\,,
\nonumber
\end{eqnarray}
where the factor $\left(-{3/\sqrt2}\right)$ comes from the 
Clebsch-Gordan coefficients relating the reduced matrix elements 
in (\ref{defM1-}) to the matrix element with the third components 
of spin and isospin ($m_s$, $m_s'$, $m_t=m_t'=\half$).
The amplitude $\mathcal{V}^\pi_{N\mathcal{R}}$ can be expressed
in terms of the elastic width of the resonance
\begin{equation}
   \Gamma_{\pi N} = 
          2\pi {\omega_0 E_N\mathcal{V}^\pi_{N\mathcal{R}}(k_0)^2
                  \over Z_{\mathcal{R}}k_0W}\,.
\label{GammapiN}
\end{equation}
Using
$
   {\rm Im} T_{\pi N\pi N} = {\Gamma_{\pi N}/\Gamma}
$
(at $W=M_{\mathcal{R}}$) 
we obtain
\begin{equation}
 {\rm Im}\,{}_pM_{1-}^{(1/2)} = -\xi_{\mathcal{R}} 
     \sqrt{k_W E^\gamma_N\Gamma_{\pi N} \over
      6\pi k_0 M_{\mathcal{R}} \Gamma^2}\; A^p_{1/2}\,,
\label{Ah2M}
\end{equation}
and similarly
\begin{equation}
 {\rm Im}\,{}_pS_{1-}^{(1/2)} = \xi_{\mathcal{R}} 
     \sqrt{k_W E^\gamma_N\Gamma_{\pi N} \over
      3\pi k_0 M_{\mathcal{R}} \Gamma^2}\; S^p_{1/2}\,.
\label{Sh2S}
\end{equation}
The above expressions differ from the standard one by
$E^\gamma_N$ appearing instead of the nucleon rest mass $M_N$.
This is a consequence of the normalization of our quark-model 
many-body state representing the recoiled nucleon which is 
normalized to 1 rather than to $M_N/E_N$. 
Adopting this convention 
requires us to slightly increase the $g_{\pi NR}$ 
coupling constant ($\sim 5$~\%) in order to reproduce 
the results with the normalization to unity.

\section{\label{sec:results}Results}

We have performed the calculation of the electro-production
amplitudes using the Cloudy Bag Model (CBM) with the same
choice of parameters as in the calculation of the scattering
amplitudes \cite{EPJ2008}.
We use the same bag radius for the excited states as for the 
ground state.
For $v(k)$ appearing in (\ref{Vmt}) we have
$$
  v(k) = r_q\,{1\over2f}\,{k^2\over\sqrt{12\pi^2\omega_k}}\,
    {\omega^0_\mathrm{MIT}\over\omega^0_\mathrm{MIT}-1}\,
            {j_1(kR)\over  kR}\,,
$$
where $r_q=1$ if $v(k$) is evaluated between the states in the 
$(1s)^3$ configuration, $r_q=r_\omega$ for the transition between 
the  $(1s)^2(2s)^1$ and the $(1s)^3$ configuration, and 
$r_q={2\over3}+r_\omega^2$ between the $(1s)^2(2s)^1$ configurations.
Here
\begin{equation}
r_\omega = 
{1\over\sqrt{3}}\,
  \biggl[\, {\omega_\mathrm{MIT}^1(\omega_\mathrm{MIT}^0-1)\over
             \omega_\mathrm{MIT}^0(\omega_\mathrm{MIT}^1-1)}\,
  \biggr]^{1/2}\,,
\label{rom}
\end{equation}
with $\omega^0_\mathrm{MIT}=2.043$ and $\omega^1_\mathrm{MIT}=5.396$.
We have adopted the conventional value of $f=76$~MeV 
which reproduces the $\pi NN$ coupling constant.
The free parameters of the model are the bag radius $R$
and the energies of the bare quark states corresponding to
the nucleon and the excited states. 
The choice of the positive sign of (\ref{rom}) fixes the relative 
sign between the quark spinors in the $1s$ and $2s$ state and 
implies $\xi_{\mathcal{R}}=+1$ in (\ref{Ah}) and (\ref{Sh}).
In particular, we have for the upper components
$u_{2s}(R)/u_{1s}(R)>0$. 

The vector mesons have not been included in our calculation
of the scattering amplitudes since their contribution turns 
out to be almost negligible in the considered energy range.
For electro-production, however, phenomenological approaches 
reveal a relatively important contribution of the $\omega$-meson 
already at lower energies.
We have therefore included the phenomenological form
of its contribution to the $K$ matrix in the elastic channel
in the form 
$$
{}_pM^{(1/2)}_{1-}(\omega\hbox{-meson})
= {1\over3}\; {M_N\over 4\pi Wm_\pi}\,
{ g_{\gamma\pi\omega}g_{\omega 1}\,k_\gamma k_\pi\,
      \rho_\omega(k_\omega)     \over 
  m_\omega^2 -m_\pi^2 + 2k_\gamma\omega_\pi}\,,
$$
where the corresponding form-factor is calculated in
our model as (see e.g. \cite{Pedro2005})
$$
   \rho_\omega(|\vec{k}_\omega|) = \int \d r \, r^2 \,
         j_0(|\vec{k}_\omega| r)(u(r)^2 + v(r)^2)\,.
$$
We use $g_{\gamma\pi\omega} = 0.374 \sqrt{4\pi/137}$
while the strong coupling is less known and is usually
assumed to lie in the range $8 < g_{\omega 1} < 20$.

\begin{figure}[h!]
\hbox to\hsize{
\includegraphics[width=80mm]{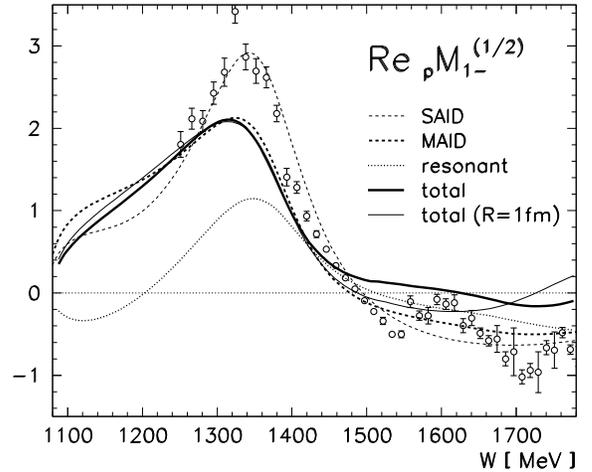}}
\caption{The real part of the $pM_{1-}^{(1/2)}$ amplitude
at $Q^2=0$ for the P11 partial wave, calculated
with $R_\mathrm{bag}=0.83\,\mathrm{fm}$ (solid thick line)
and with $R_\mathrm{bag}=1.00\,\mathrm{fm}$ (thin line).
The resonant contribution is shown separately.
The experimental points are the single-energy solutions
of the SAID partial-wave analysis \cite{SAID};
the ``SAID'' curve shows the corresponding fit;
the MAID result is from \cite{MAID}.}
\label{fig:pM1mRe}
\end{figure}

\begin{figure}[h!]
\hbox to\hsize{
\includegraphics[width=80mm]{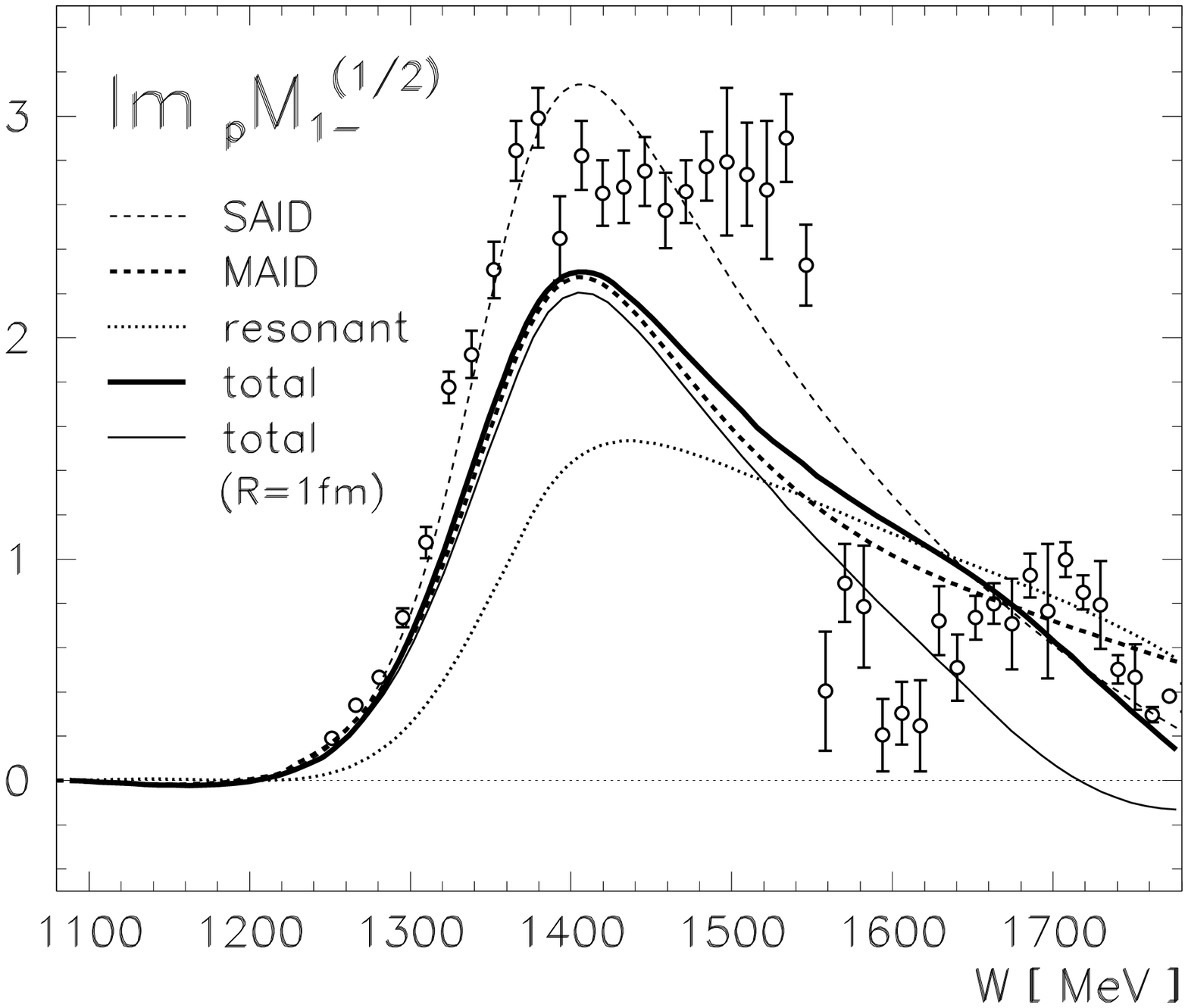}}
\caption{The imaginary part of ${}_pM_{1-}^{(1/2)}$.
Notation as in fig.~\ref{fig:pM1mRe}.} 
\label{fig:pM1mIm}
\end{figure}

\begin{figure}[h!]
\hbox to\hsize{
\includegraphics[width=80mm]{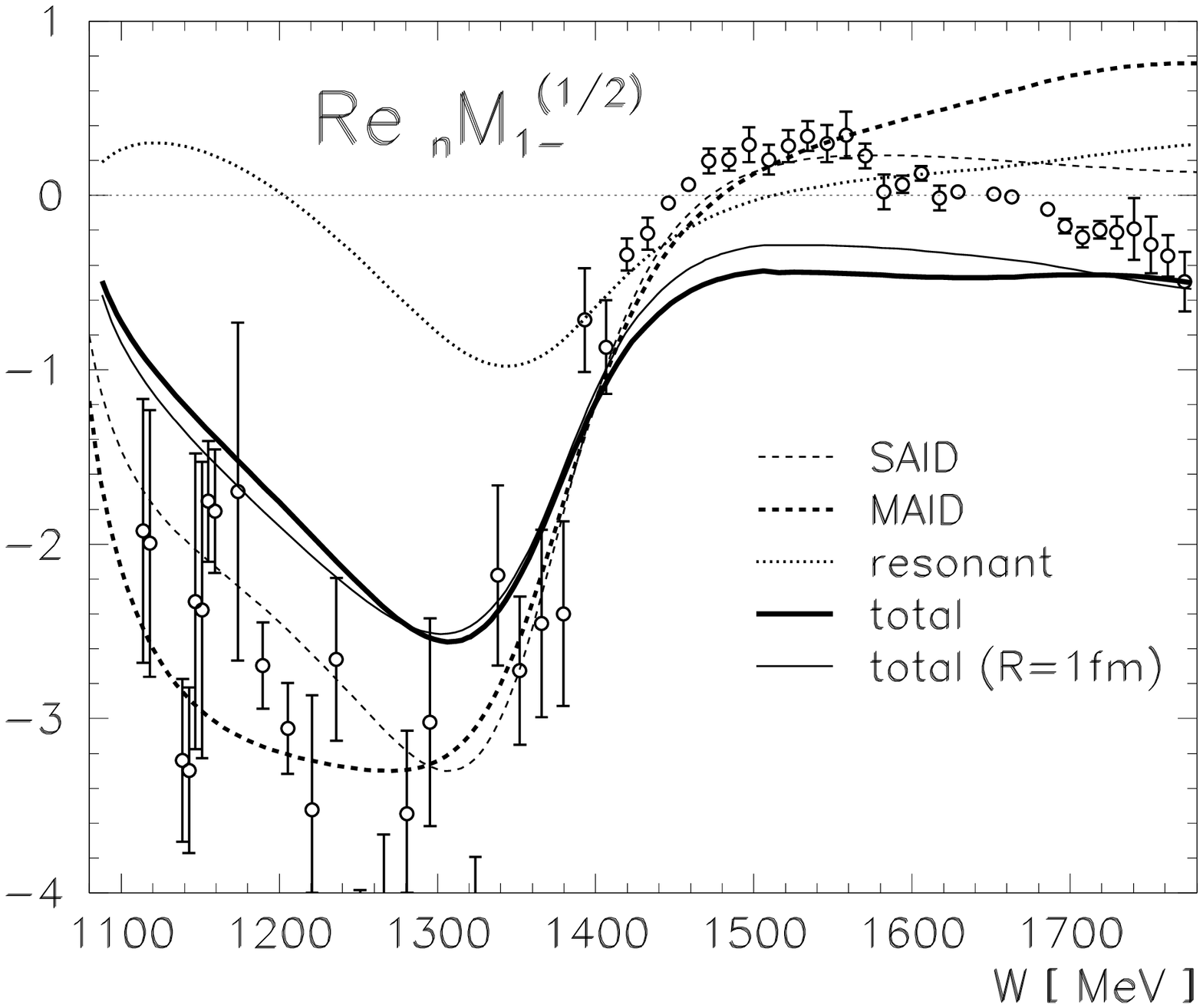}}
\caption{The real part of ${}_nM_{1-}^{(1/2)}$.
Notation as in fig.~\ref{fig:pM1mRe}.}
\label{fig:nM1mRe}
\end{figure}

\begin{figure}[h!]
\hbox to\hsize{
\includegraphics[width=80mm]{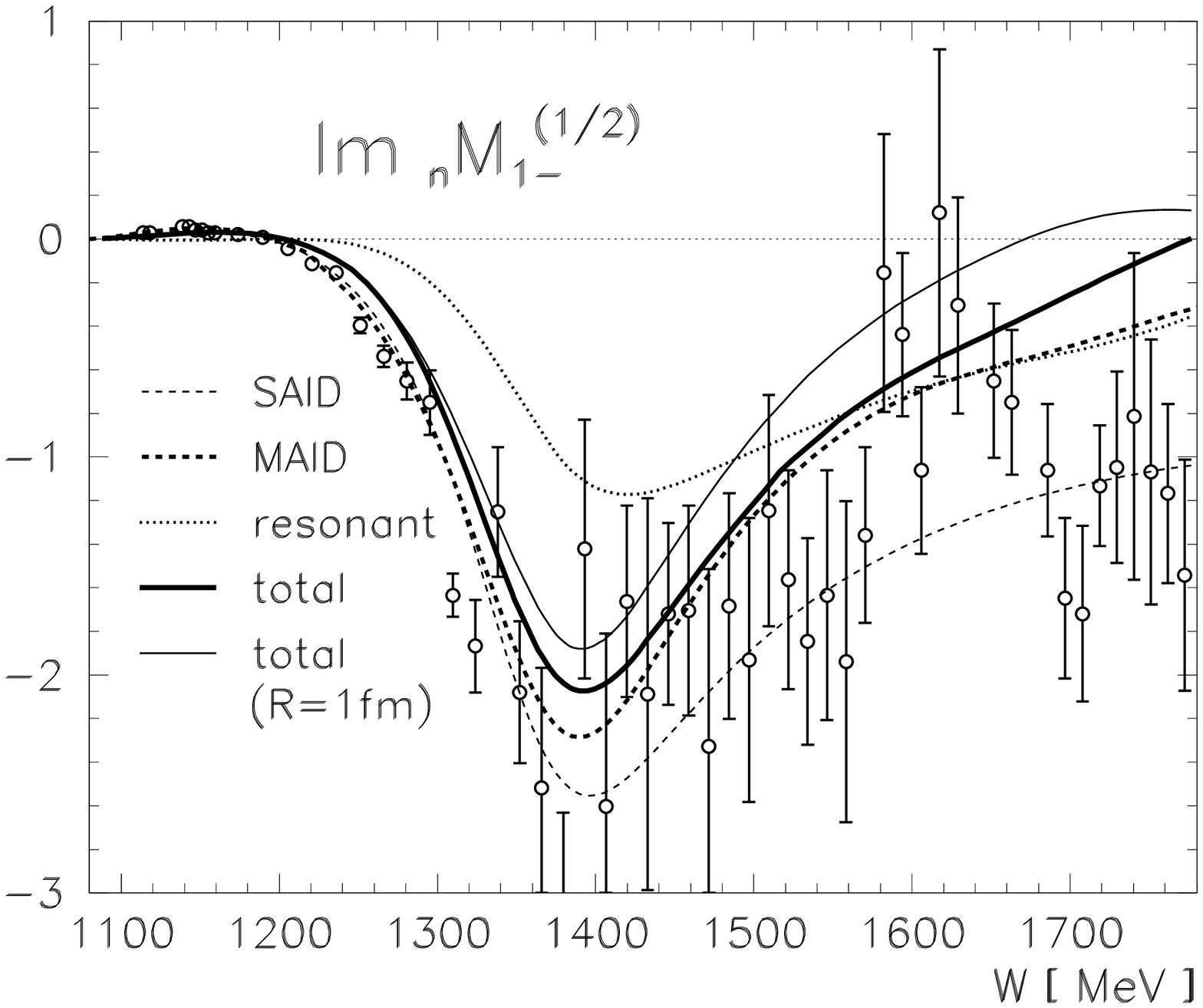}}
\caption{The imaginary part of ${}_nM_{1-}^{(1/2)}$.
Notation as in fig.~\ref{fig:pM1mRe}.}
\label{fig:nM1mIm}
\end{figure}

In figs.~\ref{fig:pM1mRe}-\ref{fig:nM1mIm} we present the 
results for the ${}_pM_{1-}^{(1/2)}$ and ${}_nM_{1-}^{(1/2)}$
amplitudes. 
Using $R=0.83$~fm and keeping the same set of model 
parameters as determined in the scattering case we reproduce 
reasonably well the experimental amplitudes.
The agreement improves if we include the
contribution of the $\omega$-meson using $g_{\omega 1}=8$.
At energies below the resonance the amplitudes are dominated 
by the background. 
This is in marked contrast to the P33 case in the region of the 
$\Delta(1232)$ which have been extensively investigated in our 
previous works (see e.g. \cite{plb1996} and \cite{epj2005}).
The photo-production amplitude in the case of the $\Delta(1232)$
is dominated by the resonant contribution and follows the shape 
of the elastic $T$ matrix in accordance with (\ref{Mres}).

\begin{figure}[h!]
\hbox to\hsize{
\includegraphics[width=80mm]{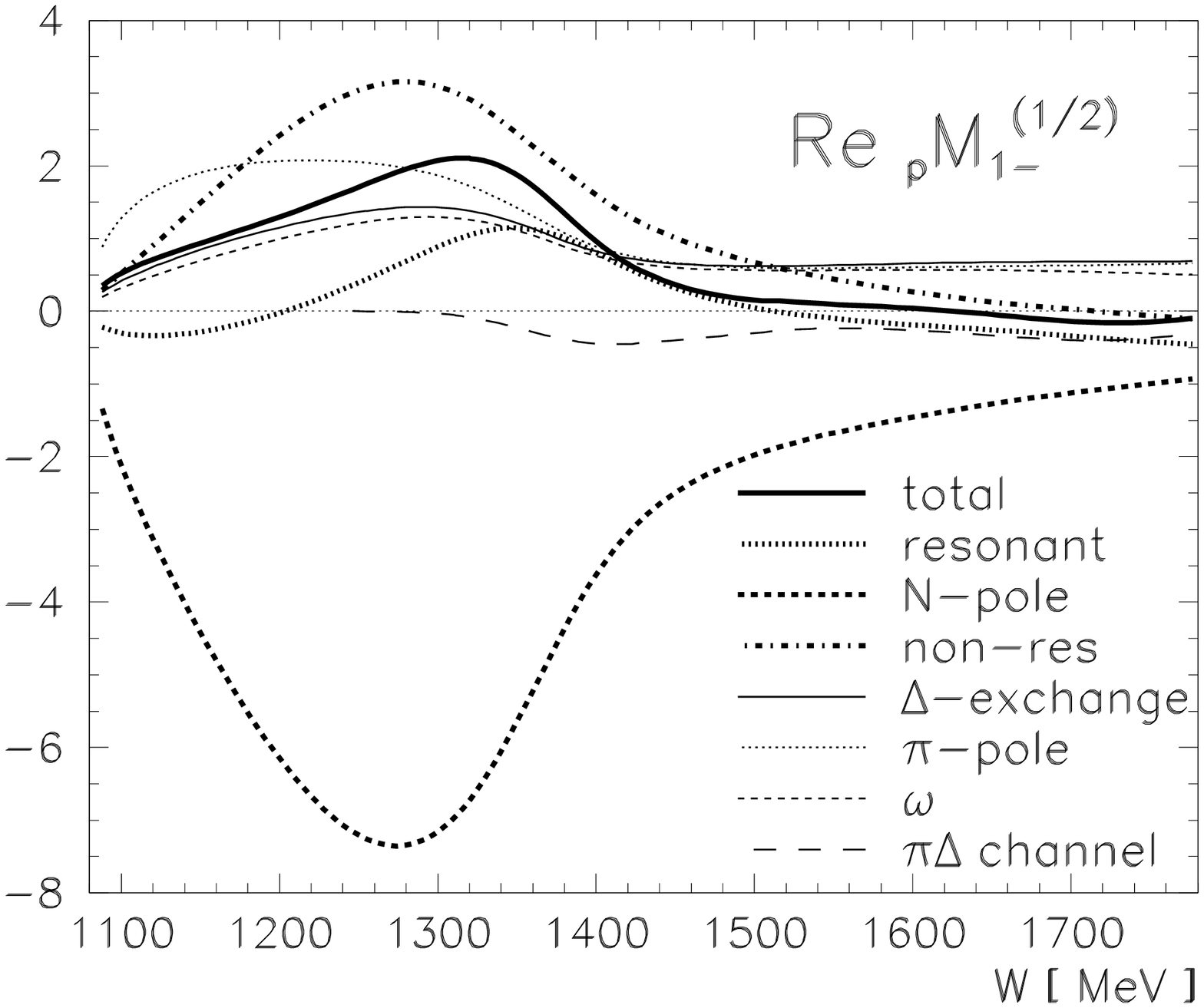}}
\caption{Contributions to $\mathrm{Re}\,{}_pM_{1-}^{(1/2)}$ 
(see text).}
\label{fig:M1mReSep}
\end{figure}

\begin{figure}[h!]
\hbox to\hsize{
\includegraphics[width=80mm]{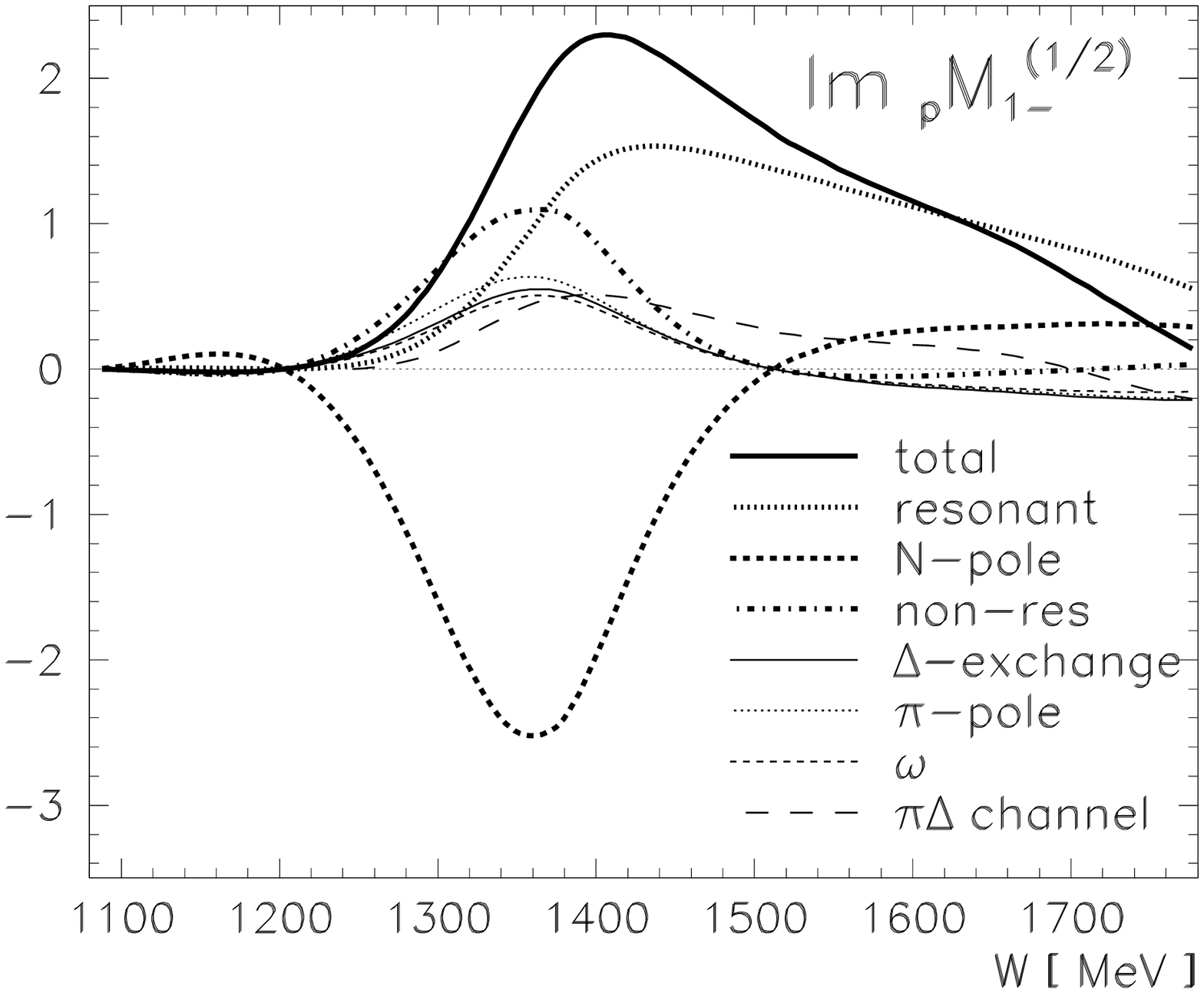}}
\caption{Contributions to $\mathrm{Im}\,{}_pM_{1-}^{(1/2)}$ 
(see text).}
\label{fig:M1mImSep}
\end{figure}

It is interesting to study different contributions to the 
total amplitude shown in 
figs.~\ref{fig:M1mReSep} and \ref{fig:M1mImSep}.
At lower energies they are dominated by 
a huge negative contribution of the nucleon pole (\ref{KHoth-res}), 
the non-resonant term (\ref{KHdir}), 
the term (\ref{KHa}) containing the u-channel isobar exchange 
(the first term of (\ref{aVgamma})) dominated by the $\Delta(1232)$, 
as well as the pion pole term (the second term in (\ref{aVgamma})).
Above the two pion threshold, the $\pi\Delta$ channel becomes
important while the contribution from the $\sigma N$ channel
turns out to be insignificant.

Taking into account the limitations and the drawbacks 
of the CBM that reproduces the static properties of the
nucleon only at the level of 10~\% to 20~\%, 
we have not considered the possibility to readjust
the model parameters in order to fit better the experiment.
We can nonetheless conclude that the model successfully 
explains the main features of the $M_{1-}$ amplitude.
The sensitivity on the variation of the bag radius (see the 
thin solid line in figs.~\ref{fig:pM1mRe}-\ref{fig:nM1mIm})
is weak except at higher energies where our model 
anyway fails to reproduce the scattering amplitudes 
above $R_{\mathrm{bag}} \sim 1$~fm.

\begin{figure}[h!]
\hbox to\hsize{
\includegraphics[width=80mm]{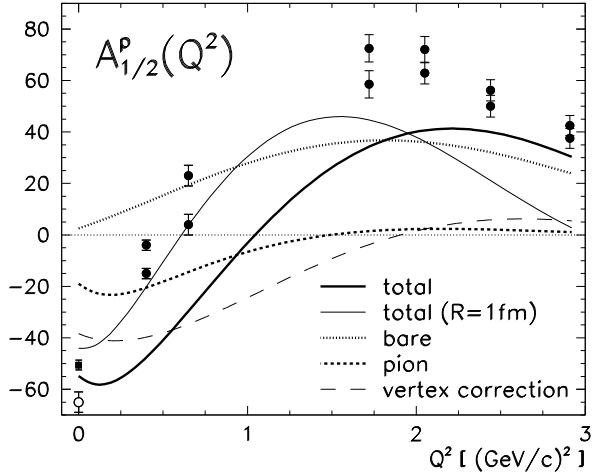}}
\caption{Helicity amplitude $A_{1/2}^p(Q^2)$ at the pole 
of the $K$ matrix ($W=1530$~MeV).  
The separate contributions include the 3q core,
the $\gamma\pi\pi'$ interaction, 
and the pion-cloud corrections to the $\gamma BB'$ vertex.  
Empty circle: PDG value \cite{PDG};
full square and circles: analyses of newer JLab experiments.
Two values at each $Q^2\ne 0$ correspond to two different
extraction approaches (see \cite{InnaCLAS} for details).}
\label{fig:Ah1520}
\end{figure}

Regarding the transverse helicity amplitude for the proton
displayed in fig.~\ref{fig:Ah1520}, we reproduce 
the value at the photon point in agreement with the 
calculation of \cite{Tiator88} within the same quark model.
This value is dominated by the pion cloud effects while the 
contribution from the bare quark core is almost negligible.
At higher $Q^2$ the quark core contribution becomes 
stronger and positive while that of the pions diminishes.
As a result the amplitude exhibits a zero crossing 
which occurs at a somewhat higher $Q^2$ than the one
extracted from the experiment.
This signifies that the pion cloud contribution may be
overestimated. Taking a larger bag radius of
$R_{\mathrm{bag}}\sim 1$~fm at which the strength
of the pion cloud becomes weaker brings
the zero-crossing value of $Q^2$ in the ballpark
of acceptable values.
A similar behaviour of the core contribution and the
pionic effects has been obtained in \cite{Dong} using
a completely different model for the quark-pion coupling.
The recent calculation in the SL model \cite{Julia-Diaz09}
also indicates a strong meson cloud contribution.

\begin{figure}[h!]
\hbox to\hsize{
\includegraphics[width=80mm]{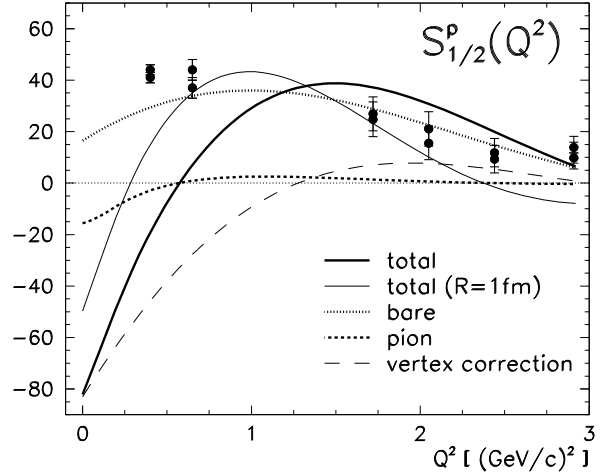}}
\caption{Helicity amplitude $S_{1/2}^p(Q^2)$. 
Notation as in fig.~\ref{fig:Ah1520}.}
\label{fig:Sh1520}
\end{figure}

\begin{figure}[h!]
\hbox to\hsize{
\includegraphics[width=80mm]{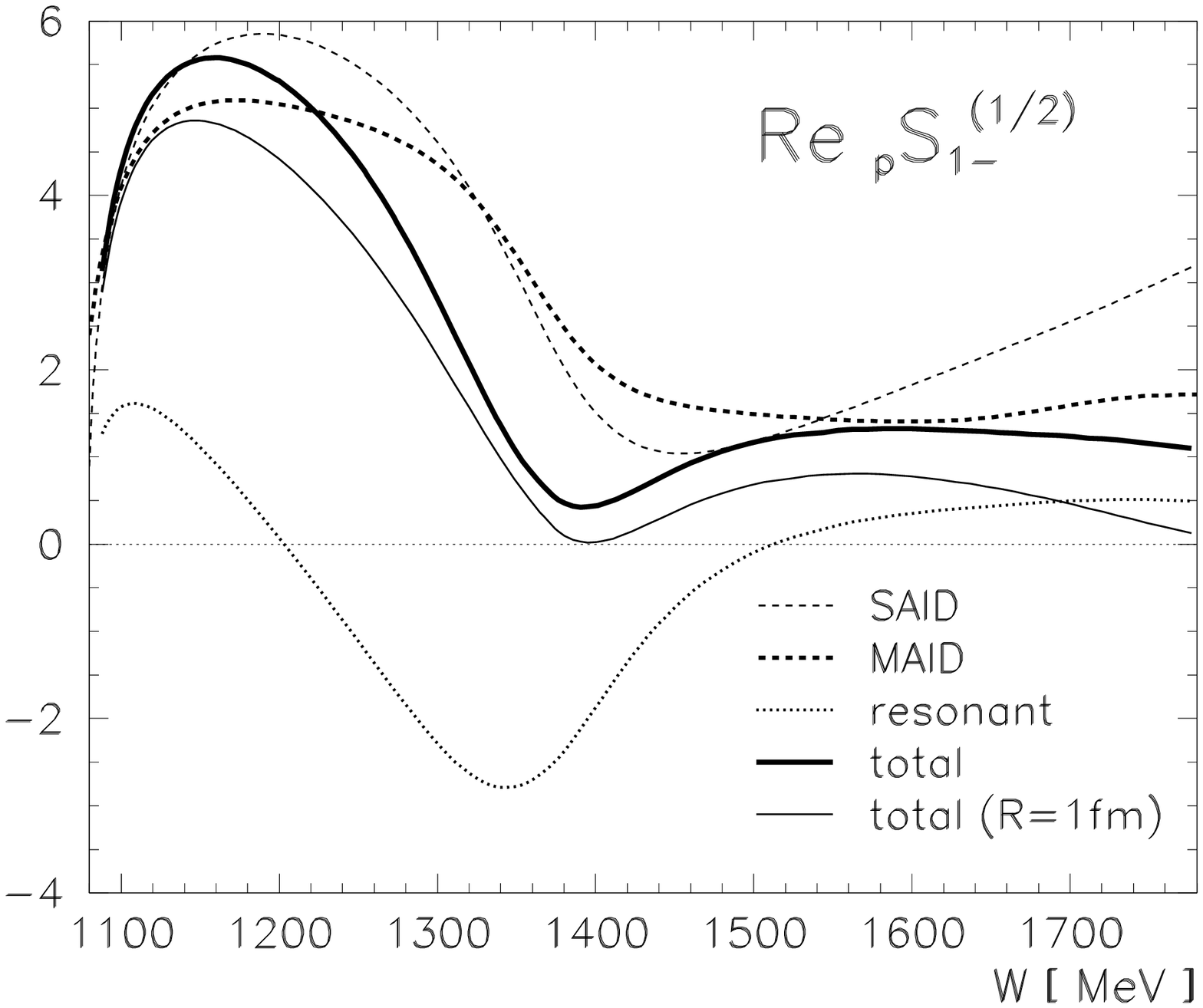}}
\caption{The real part of ${}_pS_{1-}^{(1/2)}$.
Notation as in fig.~\ref{fig:pM1mRe}.}
\label{fig:pS1mRe}
\end{figure}

\begin{figure}[h!]
\hbox to\hsize{
\includegraphics[width=80mm]{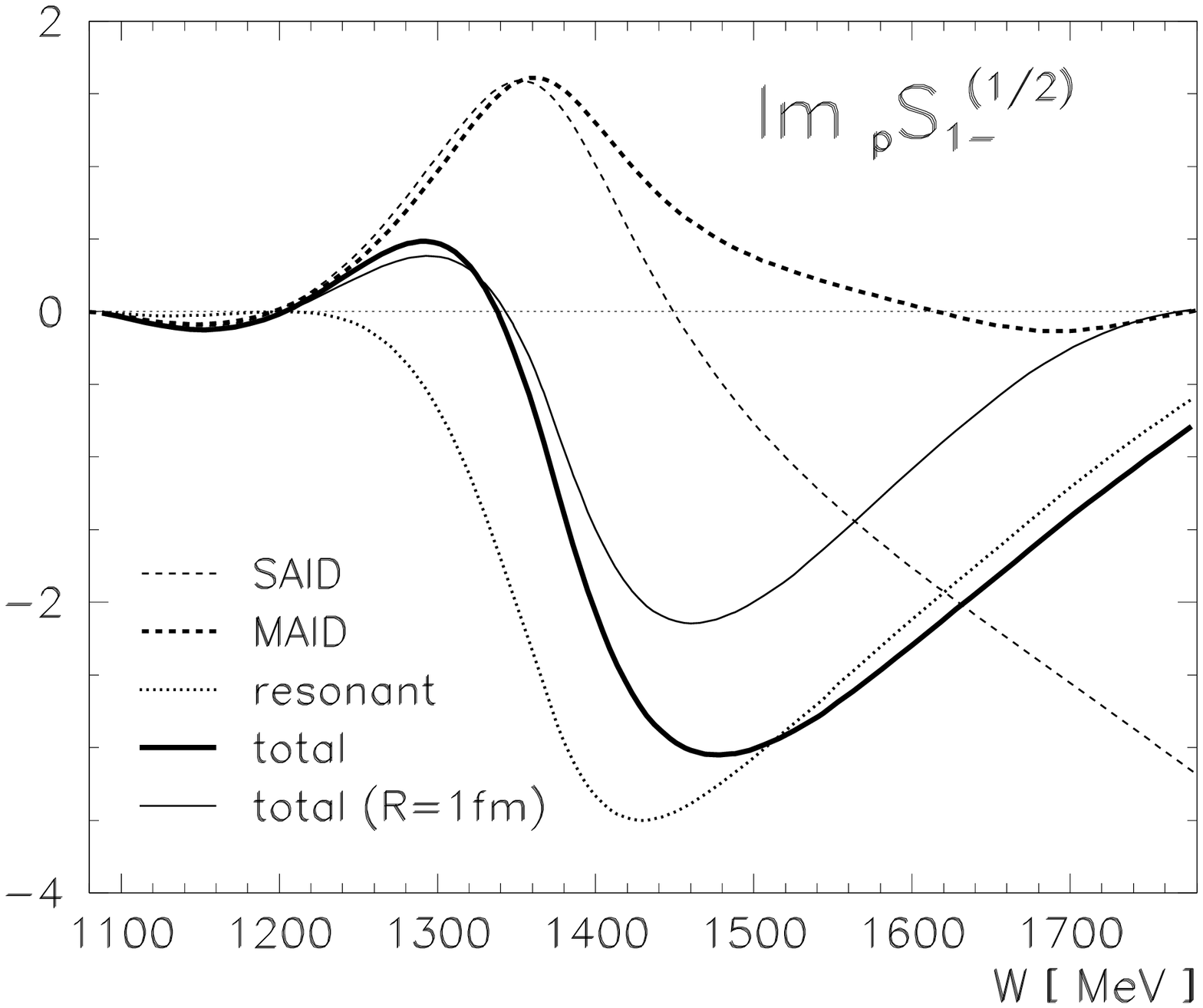}}
\caption{The imaginary part of ${}_pS_{1-}^{(1/2)}$.
Notation as in fig.~\ref{fig:pM1mRe}.}
\label{fig:pS1mIm}
\end{figure}

Such a conclusion cannot be drawn in the case of the scalar 
amplitude especially in the energy range of the resonance 
(and above), due to the rather uncertain experimental 
situation as well as the cancellation of different relatively 
large contributions to the calculated amplitude.
Since the experimental values are not available except in the 
energy range of the resonance we compare our results to the 
values deduced from the MAID and SAID analysis for  
${}_pS_{1-}^{(1/2)}$ as shown in figs.~\ref{fig:pS1mRe} 
and \ref{fig:pS1mIm}.
For $Q^2\rightarrow 0$, a reasonably good agreement
is obtained only below $\sim 1300$~MeV; above,
the imaginary part of our amplitude crosses zero much sooner 
than those of the phenomenological models and reaches a 
relatively large negative value at the resonant energy.
This is due to the effects of the pion cloud which are large 
and have the opposite sign with respect to the contribution 
from the quark core which is small at $Q^2\rightarrow 0$.
In fact, this is the consequence of the same mechanism
that governs the behaviour of the magnetic helicity
amplitude which can be seen by comparing the 
helicity amplitude $S_{1/2}$ in fig.~\ref{fig:Sh1520} to 
$A_{1/2}$ in  fig.~\ref{fig:Ah1520}.
A similar pattern for the two contributions has been found 
in \cite{Dong} but with a substantially weaker pion cloud
contribution which does not yield the zero crossing of $S_{1/2}$.
Note that the phenomenological analysis from SAID
also yields a negative value for the scalar pion production 
amplitude at $Q^2\rightarrow 0$ in the energy region of the
resonance, while the values from MAID \cite{MAID} remain
small and positive.

\section{Conclusions}

Compared with the analysis of the scattering amplitudes
in our previous work \cite{EPJ2008}, the study of 
electro-production amplitudes offers further insight 
in the dynamics of the underlying quark model.
Taking into account the approximate nature of the Cloudy 
Bag Model and the fact that we have not included  any new 
free parameters (except for the strong $\omega$ vertex), 
we have been able to reproduce surprisingly well the main 
features of the $M_{1-}$ electro-production amplitude in 
the energy range from the threshold up to $W\sim 1700$~MeV
and  for $Q^2$ up to $\sim 3$~GeV$^2$/c$^2$.
Our investigation has pointed out the important -- and in
several cases the dominant -- role played by the pion cloud,
especially in the region of low $Q^2$, being gradually
overwhelmed by the dynamics of the quark core as we go towards 
higher $Q^2$, supporting the picture in which the pion cloud 
dictates the long-range while the quark core the short-range
physics of the baryon \cite{Burkert08}.

For the reasons discussed in the previous section we are
not able to assess the quality of our prediction in the
case of the scalar amplitude.
Yet, our approach gives a rather definitive prediction
for the behaviour of the $S_{1/2}$ helicity amplitude for 
$Q^2\rightarrow 0$ which is expected to become small
or even negative in this limit.

Though we have used a relatively simple model to obtain 
the results in a particular partial wave, the method is 
applicable to a broad class of models.
Application of the method using more sophisticated models 
for the quark-meson dynamics could --  
when tested with more selective data coming from planned 
double-polarization experiments at MAMI and Jefferson Lab 
-- finally lead to the solution of the Roper puzzle.

\begin{acknowledgement}
One of the authors (S. \v{S}.)
would like to express his thanks for helpful discussions
with Inna Aznauryan and Lothar Tiator.
\end{acknowledgement}

\end{document}